\documentstyle[prl,aps,epsfig]{revtex}

\newcommand{\beq}{\begin{eqnarray}}
\newcommand{\eeq}{\end{eqnarray}}

\newcommand{\be}{\begin{equation}}
\newcommand{\ee}{\end{equation}}
\newcommand{\lwrsim}{\raise0.3ex\hbox{$<$\kern-0.75em\raise-1.1ex\hbox{$\sim$}}}

\def\C2#1#2{({\cal C}_2)_{#1}^{#2}}

\def\eq#1{Eq.(\ref{#1})}
\def\fig#1{Fig.\ref{#1}}
\def\tab#1{Tab.\ref{#1}}

\def\npb#1#2#3{Nucl.\ Phys.\ {\bf B#1} (#2) #3}

\begin{document}
\setcounter{page}{1}

\title{Instantons and $\langle A^2\rangle$ Condensate}
\author{\bf  Ph. Boucaud$^1$, J.P. Leroy$^1$, A. Le Yaouanc$^1$, J. Micheli$^1$,
O. P\`ene$^1$, F. De Soto$^2$, A. Donini$^3$, \\ H. Moutarde$^4$, J. Rodr\'{\i}guez--Quintero$^5$}
\address{Laboratoire de Physique Th\'eorique, Universit\'e de Paris XI, B\^atiment 210, 
91405 Orsay Cedex, France\\
$^2$ Dpto. de F\'{\i}sica At\'omica, Molecular y Nuclear,
Universidad de Sevilla, Apdo. 1065, 41080 Sevilla, Spain \\   
$^3$ I.N.F.N., Roma I and Dip. Fisica, Universit\`a di Roma
``La Sapienza'', P.le A. Moro 2, 00185 Rome, Italy \\
$^4$ Centre de Physique Th\'eorique Ecole Polytechnique, 
91128 Palaiseau Cedex, France \\
$^5$ Dpto. de F\'{\i}sica Aplicada,
E.P.S. La R\'abida, Universidad de Huelva, 21819 Palos de la fra., Spain}

\maketitle


\begin{flushright}
LPT-ORSAY 02-06\\
ROMA-1331-02\\
CPhT-RR-021.0202\\
UHU-FT/01-01\\
\end{flushright}

\begin{abstract}

We argue that the $<A_{\rm OPE}^2>$ condensate found in the Landau
gauge on lattices, when an Operator Product Expansion
of Green functions is performed, might be explained by instantons.
We use cooling to estimate the instanton contribution
and extrapolate back the result to the thermalised 
configuration. The resulting $<A_{\rm inst}^2>$
is similar to $<A_{\rm OPE}^2>$.

\end{abstract}

\section{Introduction}
\label{sec:intro}

Lattice calculations of the  gluon propagator and three-point Green
function  in the Landau  gauge indicate that the expected perturbative
behaviour at  large momentum $p$ has to be corrected by a 
$O(1/p^2)$ contribution  sizeable up to 10 GeV \cite{football,Boucaud:2000nd,Boucaud:2001st,Becirevic:1999hj,desoto}. 
An understanding of this contribution as
the effect of an $A^2 \equiv A_\mu^a A_a^\mu $  condensate (in the
Landau gauge $A^2$ is the only mass dimension-two operator liable to have a
v.e.v.)  has been  gained by verifying that 
two independent Green functions could be described by the perturbative
contribution corrected by the effect of one common value of $<A^2_{\rm OPE}>$,
as expected from OPE. The physical origin of this condensate is an important question, possibly involving  
the non-trivial topology of the QCD vacuum. In particular, 
 instantons provide an interesting insight into a wide 
range of low energy QCD properties (\cite{shuryak} and refs therein).  
They have been  put into evidence on
the lattice using different cooling procedures. In this letter we 
claim that instantons provide for $<A^2>$ a value close 
to what  is needed for the OPE fit to Green functions. 

We propose a method to identify instantons from the cooled gauge
configuration, count them and measure their radii; we also check
that these results are compatible with the  instanton number
deduced from the two-point correlation function of an instanton. We
then estimate $<A_{\rm inst}^2>$, the contribution  of the
instantons to $<A^2>$ in cooled  configurations, extrapolate back
to the thermalised configurations (zero cooling sweeps) and finally,
compare the outcome with  the OPE estimate.

\section{Cooling and instanton counting by shape recognition}
\protect\label{sec:cool}

\subsection{cooling}

In order to study the influence of the underlying classical properties 
of a given lattice configuration, the first step will be to isolate these
structures from UV modes. The method we use is due to Teper  \cite{teper};
it  consists in replacing each link by a unitary matrix proportional to
the staple.  A {\it cooling sweep} is performed after replacing  all the
links of the lattice.
This procedure introduces largely discussed biases, such
as UV instanton disappearance and instanton--anti-instanton pair
annihilations, that  increase with the number of cooling sweeps;
alternative cooling methods have been proposed 
(see for example \cite{garciaperez} and refs. therein) 
to cure these diseases. We will try to reduce them by identifying 
the instantons after a few cooling sweeps and extrapolating  back to the
thermalised gauge configuration.

\subsection{Instantons}
\label{inst}

Instantons (anti-instantons) are classical solutions \cite{thooft} of the
equations of motion. We work in the Landau gauge which is defined on the lattice by
minimizing $\sum_x A_\mu(x)^2$. For an instanton solution this prescription leads
to  the  singular Landau gauge, where the gauge field is\cite{thooft} 

\beq
A^{(I) a}_\mu = \frac{2 \overline{\eta}^{a}_{\mu \nu} x_\nu \rho^2}{x^2 \left(x^2+\rho^2\right)} \ ,
\label{field}
\eeq
\noindent  $\rho$ being the instanton radius. The instanton 
has been chosen centered at the origin with a conventional color orientation.
These solutions have $Q=\pm 1$ topological charge, 

\beq
Q \equiv \frac{g^2}{32\pi^2} \int d^4x F^{\mu\nu}_a \widetilde{F}^{a}_{\mu\nu}
= \pm  \frac{g^2}{32\pi^2} \int d^4x F^{\mu\nu}_a {F}^{a}_{\mu\nu} \ , 
\label{Q}\eeq

\noindent where  $\widetilde{F}_{\mu\nu}=\frac{1}{2}\epsilon_{\mu\nu\rho\sigma}
F^{\rho\sigma}$. 
From eq. (\ref{field}) the topological charge density is:

\beq
Q_\rho(x)=\pm\frac{6}{\pi^2\rho^4}\left(\frac{\rho^2}{x^2+\rho^2}\right)^4\ ,
\label{ins}
\eeq

On the lattice, the topological charge density will be computed as

\beq
Q_{\rm latt}(x)\ =\ \frac{1}{2^9\pi^2} \sum_{\pm 1}^{\pm 4} \widetilde{\epsilon}_
{\mu\nu\rho\sigma} Tr[\Pi_{\mu\nu}(x)\Pi_{\rho\sigma}(x)]\ ,
\label{qlat}
\eeq

\noindent with $\Pi_{\mu\nu}(x)=U_\mu(x)U_\nu(x+\mu\hat{a})U^\dagger_
\mu(x+\mu\hat{a}+\nu\hat{a})U^\dagger_\nu(x+\nu\hat{a})$, and 
$\widetilde{\epsilon}_{\mu\nu\rho\sigma}$ the antisymmetric tensor, with an
extra minus sign for each negative index.

\subsection{identification of instantons}
\label{identification}

A common belief is that an instanton liquid  gives a fair description of
important features of the QCD vacuum.  Along this line   we will try a desciption
of our cooled gauge configuration as  an  ensemble of non-interacting instantons
with random positions and color orientations. We hence also neglect the 
interaction-induced instanton deformations and correlations.    Although the
instanton ensemble for the QCD vacuum cannot be considered as  a dilute
gas~\cite{shuryak,teper,garciaperez,thooft,diakonov,verb,hutter}, this crude assumption 
allows a qualitatively reasonable picture, especially near the instanton center.

Many enlightening works have studied the instanton properties from lattice gauge
configurations\footnote{A detailed comparison of our results to their's will
appear in a forthcoming extensive work.}, among which \cite{chu,deForcrand:1997sq,smith,michael}. As for 
us, we start by  searching regions where the topological charge density looks
like that of \eq{ins}. Starting from each local maximum or minimum of $Q_{\rm
latt}(x)$  we integrate over all neighbouring points with $|Q_{\rm latt}(x)| \ge
\alpha  |Q_{\rm latt}(x_{max})|$, for different values of $\alpha$ ranging  from
$0.8$ to $0.4$. A local extremum is accepted to be an (anti-)instanton  if the
ratio $\epsilon$ between the lattice integral and its theoretical  counterpart,
$Q_\rho (x)$, 

\beq
\epsilon\ =\ \left( 1-3\alpha^{1/2}+2\alpha^{3/4}\right)^{-1}  \ 
\int_{{x}/\ |Q({x})|\ge\alpha |Q(0)|}d^4x\ Q_{\rm latt}(x) 
\label{epsilon}
\eeq

\noindent shows a  plateau when $\alpha$ is varied.  
Indeed for a theoretical (anti-)instanton $\epsilon=1$ ($\epsilon=-1$)
for any $\alpha \in [0,1]$. 
As a cross-check of self-duality, this instanton shape recognition (ISR)
 procedure is applied on the lattice to both expressions for $Q$
 introduced in eq. (\ref{Q}).

\subsection{Instanton numbers and radii} 

The ISR method  resolves the semiclassical structure on the lattice
with only $\sim 5$ cooling sweeps. This early recognition reduces
the possible cooling-induced bias.

Whenever this method succeeds,  we measure the radius of the accepted 
instanton  in two separate ways: from the maximum value of $Q$, written
$Q_{\rm latt}^{\rm max}$,
at the center of the instanton, and from the number 
$N_\alpha$ of lattice points in the volume
$\int_{{x}/\ |Q({x})|\ge\alpha |Q(0)|}d^4x$.  
We find:

\beq
\rho/a\ =\sqrt[4]{\frac 6 {\pi^2 Q_{\rm latt}^{\rm max}}} =  
\frac{1}{\sqrt{\alpha^{-1/4}-1}} \sqrt[4]{\frac{2N_{\alpha}}{\pi^2}}\ ,
\eeq

\noindent 
These two measures of the radius agree within $10\%$. 

\section{gluon propagator} 
\label{sec:prop}

\subsection{The instanton gauge-field correlation function}

The classical gauge-field two-point correlation function verifies,
 for any position and color orientation,

\beq \frac 1 8 \sum_a\overline G_{\mu \nu}^{a a}\ 
\equiv  \frac 1 {8V} \sum_a\left(A^{(I) a}_\mu(k) A^{(I) a}_\nu(-k)\right)
  \ = \overline G^{(2)}(k^2) \ \left( \delta_{\mu \nu}-\frac{k_\mu
k_\nu}{k^2} \right) \   \ ,
\label{tensor-prop} \eeq

\noindent where $V$ is the volume in the euclidean four-dimensional space and
$A^{(I) a}_\mu(k)$ is the Fourier Transform of \eq{field}. The 
resulting scalar form factor is:

\beq \overline G^{(2)}(k^2) \ = \ \frac{32 \pi^4}{V k^6} \left( 1 - \frac{(k
\rho)^2}{2} K_2(k \rho) \right)^2 \ , 
\label{G2} \eeq

\noindent  $K_2$ being a Bessel 
function~\cite{Gradshteyn}. Eq. (\ref{G2}) equally applies 
to instantons and anti-instantons\footnote{A similar analysis is being
parallelly performed, in other context, by Broniowski and Dorokhov\cite{Bron02}}.
 
In a perfect gas approximation for an ensemble of $n_I$ ($n_A$) 
(anti-)instantons of radius $\rho$, the classical gauge-field correlation
function is simply given by \eq{G2} times the number of instantons and
anti-instantons, $n_I+n_A$. This correlation function is the contribution of the
background field to the gluon propagator. We expect this formula to describe the
behaviour of the lattice gluon propagator once the  effect of quantum UV fluctuations is removed
by the cooling procedure~\cite{teper}.   The effect of instanton interactions  is
known~\cite{diakonov,verb} to modify the instanton shape far from its center, in
the IR region.   But the large $k^2$ behaviour should be  appropriately  given by
\eq{G2} i.e. $\propto 1/k^6$. This is shown in \fig{Fig1}.a for one 
generic lattice gauge field configuration.

The theoretical lines in that plot  \footnote{We multiply by $k^2$ to compute a
dimensionless object  and perform the matching.} are generated by \eq{G2} using
the average radius $\rho$ and   $n_I+n_A$ computed from the ISR method. Note that
the  matching improves with the number of cooling sweeps. This agrees with the
expectation that decreasing the instanton density reduces the instanton 
deformation and that quantum fluctuations are damped by cooling.
Reversely, if we know the average radius from the ISR method, we can 
compute $n_I+n_A$  from the fit to the measured propagator.

\subsection{The hard gluon propagator}

Let us now consider a hard gluon of momentum $p_\mu$ propagating
in an instanton gas background. The gluon interacts with the
instanton gauge field.  This can be computed with Feynman graphs
  and  it is easy  to see that  when the instanton modes $k_\mu$ verifies $k_\mu  \sim 1/\rho 
\ll p_\mu$, the dominant contribution is an $O(1/p^2)$ correction
to the perturbative gluon. This correction is equal to the
standard OPE Wilson coefficient \cite{Boucaud:2000nd,Boucaud:2001st}
 times $<A^2_{\rm inst}(0)>\equiv 1/V\int
d^4 x \left( A^{(I)} \right)^2(x)$. We will now proceed to estimate this instanton-induced 
condensate.

\section {$<A^2>$ condensate}

\subsection{$<A^2>$  in instantons}

From \eq{field} we get
 
\beq 
<A_{\rm inst}^2(n_c)> \equiv  \frac{n_I+n_A}{V} \ 
\int d^4x \sum_{\mu,a} A^{(I) a}_\mu A^{(I) a}_\mu
\ = \ 12 \pi^2 \rho^2 \frac{n_I+n_A}{V} \ .
\label{E9}
\eeq
where $\rho$ is the average instanton radius in the considered 
cooled configuration and $n_c$ is the number of cooling sweeps.

\begin{figure}[t!]
\begin{center}
\begin{tabular}{cc}
\hspace{-0.5cm} \epsfxsize7.1cm\epsffile{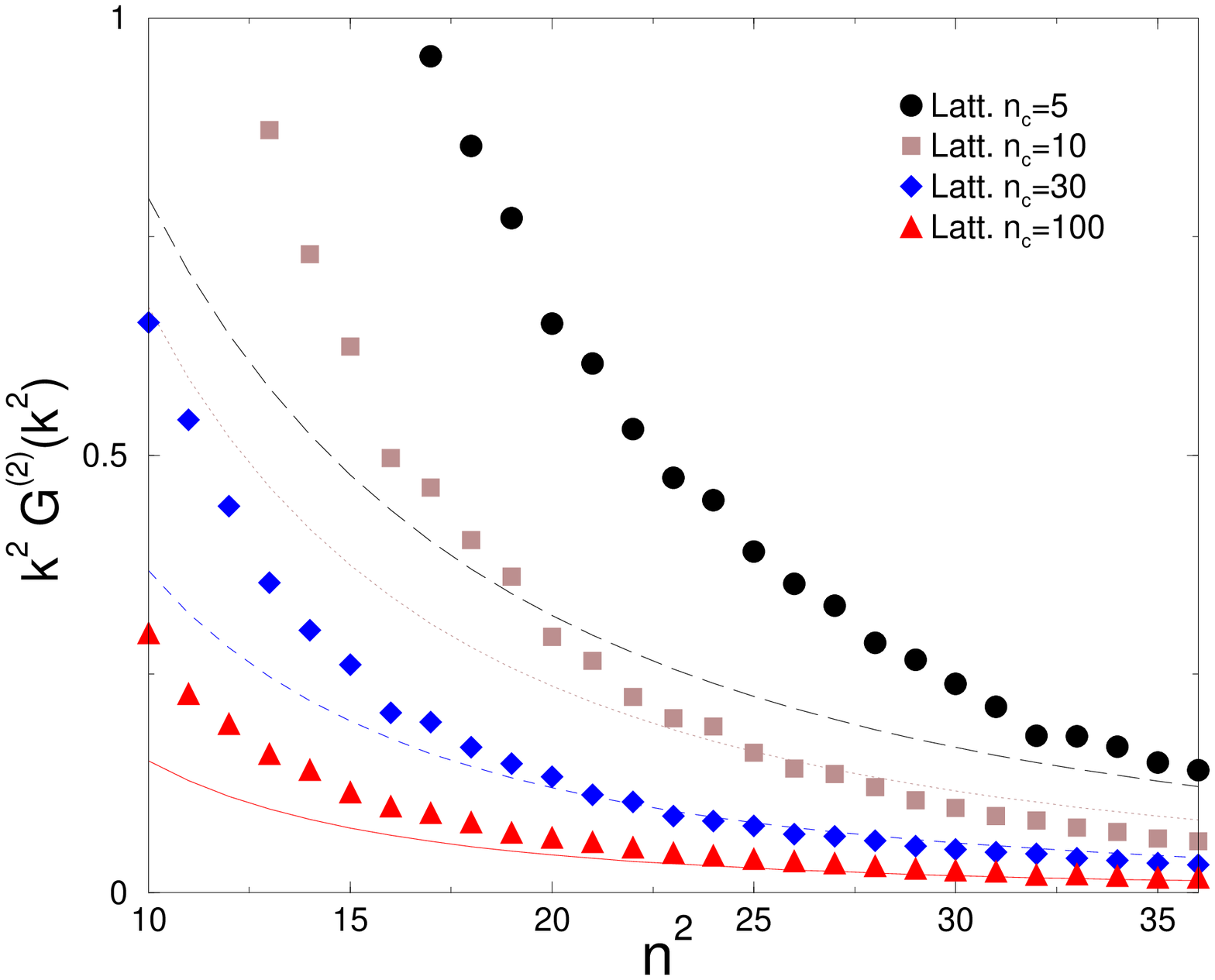} & 
\hspace{0.5cm} \epsfxsize7.6cm\epsffile{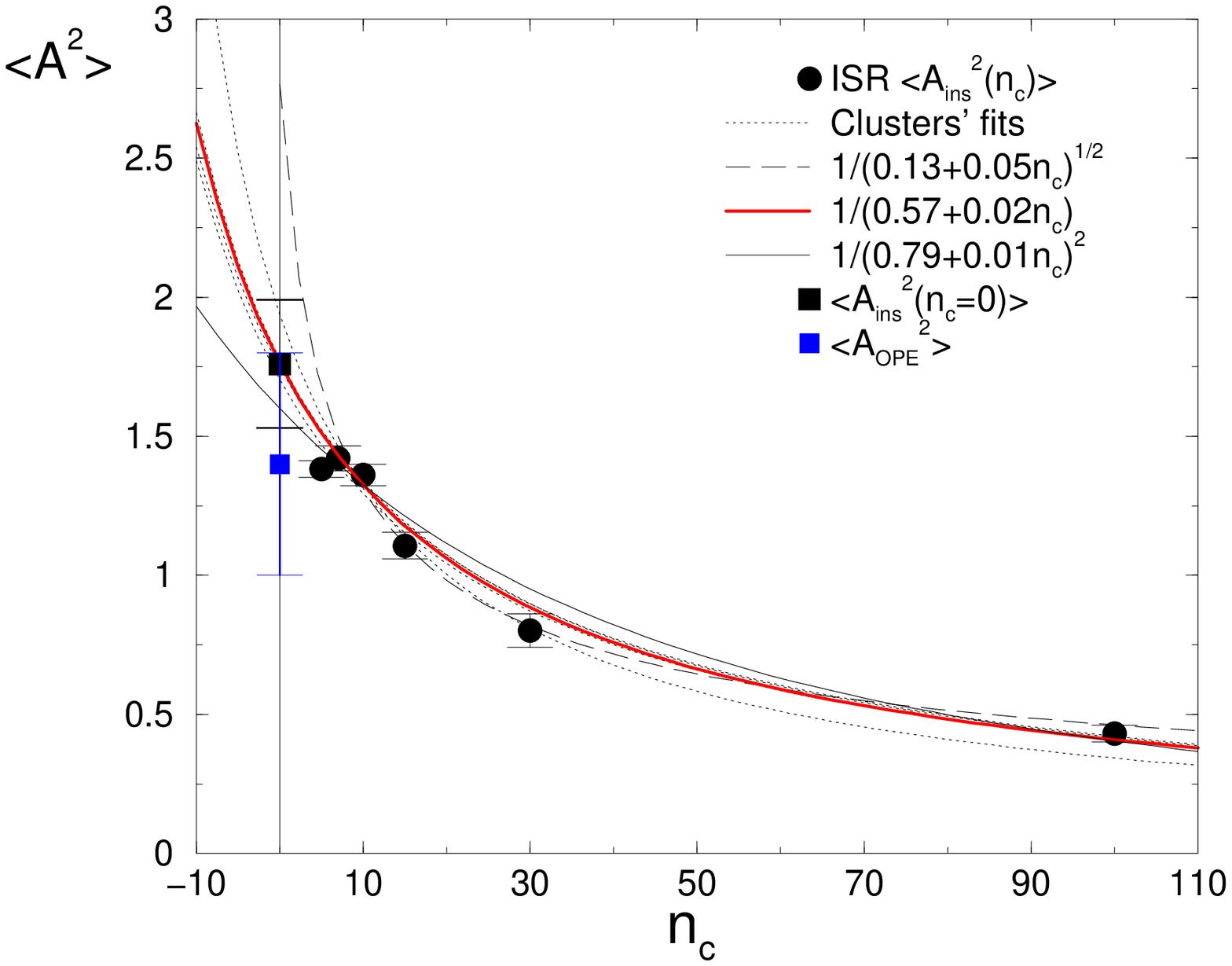} 
\end{tabular}
\caption{\small 
In fig. (a) (left) we present the lattice gluon propagators after several 
number of cooling sweeps (points) and the corresponding theoretical 
instanton gauge-field correlation functions (lines) in
the perfect instanton gas approximation, \eq{G2}, plotted as a function 
of $n^2$ ($n_\mu \equiv  L/(2\pi)\ k_\mu$, $L$ being the lattice length).
In fig. (b) (right), we show the extrapolation at zero cooling of $<A^2 (n_c)>$ 
in physical units (for three different trial functions).} 
\label{Fig1}
\end{center}
\end{figure}

 We use an ensemble of 10 independent gauge configurations\footnote{Considering
 the present size of our systematic uncertainties we  did not consider it
 worthwhile to increase further the statistics.} at $\beta=6.0$ on a $24^4$
 lattice. Each configuration  has been cooled and after 5,7,10,15,30 and 100
 cooling sweeps transformed into the Landau gauge. Using the ISR method on each
 gauge configuration, we obtain the results of \tab{Tab1}. In this table we also 
present the number of (anti-)instantons  and the
corresponding value for $<A^2(n_c)>$ computed by a Correlation  Function Fit
(CFF) i.e. a fit of the lattice propagators to the instanton  correlation
function, \eq{G2}. The CFF method is expected to be affected differently from the
ISR method  by systematic uncertainties: instanton interactions, deformations 
and  quantum fluctuations (as we can see in \fig{Fig1}.a,  at low momentum), and
we therefore consider as quite encouraging the  qualitative agreement - becoming
quantitative at large $n_c$ - between ISR and CFF results. We have, for
simplicity, translated our lattice results into physical units using, for all
values of $n_c$, the $n_c=0$ inverse lattice  spacing, $a^{-1}(n_c = 0) = 1.996$
GeV (at $\beta=6.0$). This simple recipe overlooks the effect of cooling on the
lattice spacing  (see ref. \cite{smith} and refs. therein) but this
simplification becomes harmless after extrapolating back our results to $n_c =
0$.

\subsection{$<A_{\rm inst}^2 (n_c)>$ at zero cooling.}

The instanton number depends on the number of cooling sweeps.  This result may
imply that the cooling procedure destroys not only  quantum UV fluctuations but
something else from the semiclassical background  of gauge fields. To lessen this
problem we take advantage of the  early recognition of the instanton content in a
gauge configuration ensured by the ISR method and perform an extrapolation
\cite{negele} to $n_c = 0$  of the ISR results for $<A^2_{\rm inst} (n_c)>$ in
the table.  We then obtain  (see fig.1.b): 

\beq 
<A^2_{\rm inst}(n_c = 0)> = 1.76(23) \; {\rm GeV}^2 \ .
\label{a2inst}
\eeq

We have used a form $a/(b+n_c)$  to fit and extrapolate. We have also varied
a little this functional form to check the stability of the extrapolation. 
We take this result as indicative of the non-perturbative instanton 
contribution to the $<A^2>$ condensate.
If we applied other lattice estimates of instanton gas parameters taken from 
the available literature to \eq{E9}, the value of $<A^{(I)^2}>$ would range~\footnote{We
  use the parameters obtained in \cite{DeFor} for simulations on
  different lattices and $\beta$'s with a cooling improved to let
  scale invariant instantons solutions exist for large
  enough instanton sizes. We only quote anyway the results where the
  packing rate, $\pi^2/2 (\rho/L)^4 (n_I + n_A)$, is as much as 1, since our  
  method to estimate $<A^2_{\rm inst}(n_c)>$ assumes limited overlap between instantons.} from 
  1 to 2 GeV$^2$. 
On the other hand, parameters from
  instanton liquid based phenomenology\cite{shuryak}
  yield estimates
  of the order of 0.5 GeV$^2$.
As the quoted error in \eq{a2inst} is only statistical, this
last range somehow estimates a certain systematic uncertainty.

\begin{table}[b]
\begin{center}
\begin{tabular}{|c|c|c|c|}
 $n_c$  & $ \rho$ (fm) & $n_I+n_A$ & $ <A^2>$ \ (GeV$^2$) \\
\cline{2-4}
 \begin{tabular}{c} 
 \\ 
 \cline{1-1} 
 5 \\ 7 \\ 10 \\ 15 \\ 30 \\ 100  
 \end{tabular}   
 &
\begin{tabular}{c} 
ISR \\ 
\cline{1-1}
           0.329(2) \\ 
           0.361(2) \\ 
           0.394(4) \\   
           0.417(5) \\
           0.452(9) \\
           0.53(1)
\end{tabular} 
& 
\begin{tabular}{c|c}
ISR & CFF \\ 
\cline{1-2}
          87(2) & 93(10) \\ 
          74(2)& 59(1)\\ 
          60(1) & 38(1)\\   
          43(1)& 28(1) \\
          26(1) & 19(1) \\
          10(1) & 9(1)
\end{tabular} 
&
\begin{tabular}{c|c} 
ISR & CFF \\ 
\cline{1-2}
          1.38(3) &  0.9(8) \\ 
          1.42(5) & 1.12(2) \\
          1.36(4) & 0.86(2) \\  
          1.11(4) & 0.72(2) \\
          0.80(6) & 0.57(3) \\
          0.43(3) & 0.37(3)
\end{tabular} \\
\end{tabular}
\caption{Estimates of $<A^2_{\rm inst} (n_c)>$.}
\label{Tab1}
\end{center}
\end{table}


\subsection{Comparison with $<A^2>$ from OPE.}

Our instanton estimate of $\langle A^2 \rangle$ is a
semiclassical one, deprived of the necessary
UV fluctuations, and therefore not
directly comparable with~\cite{Boucaud:2000nd}  
$\langle A^2_{\rm OPE}\rangle(10 {\rm GeV})=2.4(5)~$GeV$^2$.
There is of course
no exact recipe to compare both estimates, since the separation
between the semiclassical non perturbative
domain and the perturbative one cannot be
exact.
We may appeal to the fact that at the
renormalisation point $\mu$, the radiative
corrections are minimised; therefore
a semiclassical estimate must best
correspond to $\langle A^2_{\rm OPE} \rangle$ at some reasonable
$\mu$. In the example of $\phi^2$
vacuum expectation in the spontaneously
broken $\phi^4$ model
given in \cite{SVZ}, one finds indeed
that it equals the classical estimate
for $\mu$ around the spontaneously generated
mass. In our problem,
one could guess that the corresponding
scale should typically be around $1/\rho \simeq 0.7$ GeV,
or some gluonic mass, a very low scale
anyway. We cannot run $\langle A^2_{\rm OPE}\rangle(10 {\rm GeV})$
down to such a low scale~\cite{Boucaud:2000nd},

\beq 
A^2_{R,\mu} = A^2_{R,\mu_0} \ \left( 1 + 
\frac{35}{44} \ \ln{\frac{\mu}{\mu_0}}\ / \ \ln{\frac{\mu}{\Lambda_{\rm MOM}}} \right) \ .
\label{running}
\eeq

\noindent we therefore stop arbitrarily $\mu \sim 2.6$ GeV, where the 
OPE corrected perturbative running of the Green functions fails to correctly describe their
behaviour.  This scale of 2.6 GeV  turns out to be of the same
order\cite{Boucaud:2000nd} as the critical  mass\cite{Novi} for the gluon
propagator.  At this scale, we obtain:

\beq
<A^2_{\rm OPE} (2.6 \ {\rm GeV})> = 1.4(3)(3) \; {\rm GeV}^2 \ , 
\label{a2ope} 
\eeq

\noindent where the first quoted error just propagates the uncertainty from 
the OPE determination of $<A^2>$ and the second one takes into account,
 in the way proposed in ref.~\cite{Becirevic:1999hj}, higher
orders\footnote{We re-write \eq{running} in terms of the coupling
constant renormalised in other schemes, such as $\overline{\rm MS}$.} in
$\alpha_s$ for running.

\section{Discussion and conclusions}
\protect\label{sec:conclusions}

We are aware that our method of comparison of $<A^2_{\rm inst}(n_c =
0)>$ and $<A^2_{\rm OPE}>$ suffers from a lot of arbitrariness 
and approximations (such as the perfect gas approximation,  possible errors in the
instanton identification, the uncertainty in the  extrapolation to zero cooling
sweeps, etc.). We have taken care to crosscheck  our estimates  by comparing
different methods at each step of the  computation, in particular the ISR and 
CFF (see \fig{Fig1}.a and Tab. I). A comparison with direct
``measurements'' of the $\langle A^2 \rangle$ condensate from cooled 
lattice configurations could be thought as an additional crosscheck. 
Qualitative agreement is found for a large enough number of cooling 
sweeps, but this agreement is manifestly destroyed by UV fluctuations
already for $n_c \sim 30$. Of course, by using ISR and instanton gas
approximation we sharply separate UV fluctuations from the
semiclassical background. All these imprecisions seem anyway inherent
to the subject.

 With this in mind, we nevertheless take the fair  agreement between Eqs.
(\ref{a2inst}) and (\ref{a2ope}) as a convincing indication  that  the $A^2$
condensate receives a significant instantonic contribution. 
{\it In other words, the instanton liquid picture might yield the explanation
for the $1/p^2$ corrections to the perturbative behaviour of  Green functions
computed with thermalised configurations on the lattice.}

\section*{Acknowledgements.}

These calculations were performed on the  Orsay APEmille  purchased thanks to
a funding from the Minist\`ere de l'Education Nationale and the CNRS. We are
indebted  to  Bartolome Alles, Carlos Pena and Michele Pepe for illuminating 
discussions. A. D. acknowledges the M.U.R.S.T. for financial support through
        Decreto 1833/2001 (short term visit programme), F. S. acknowledges
the Fundaci\'on C\'amara for financial support. A.D. wishes to thank 
        the LPT-Orsay for its warm hospitality. This work was supported 
        in part by the European Union Human Potential Program
under contract HPRN-CT-2000-00145, Hadrons/Lattice QCD.

\end{document}